\documentclass[11pt]{article}

\usepackage[a4paper,margin=1in]{geometry}
\usepackage{amsmath,amssymb}
\usepackage{graphicx}
\usepackage{authblk}
\usepackage{setspace}
\usepackage{cite}
\usepackage{hyperref}

\begin{document}

\begin{center}
{\LARGE \bfseries The giant anomalous Hall and Nernst effects in Kagome permanent}

\vspace{0.4cm}

{\LARGE \bfseries magnets \textit{RCo}$_5$}

\vspace{0.5cm}

{\large
Weian Guo$^{1}$, Pengyu Zheng$^{1}$, Rui Liu$^{1}$, Yiran Peng$^{1}$, Ying Yang$^{1}$ and Zhiping Yin$^{1,2*}$
}

\vspace{0.3cm}

{\large
$^{1}$School of Physics \& Astronomy and Center for Advanced Quantum Studies, Beijing Normal University, Beijing 100875, China
}

\vspace{0.15cm}

{\large
$^{2}$Key Laboratory of Multiscale Spin Physics (Ministry of Education), Beijing Normal University, Beijing 100875, China
}

\vspace{0.15cm}

{\large
* Contact authors: yinzhiping@bnu.edu.cn
}
\end{center}

\vspace{0.6cm}

{\Large \bfseries Abstract}

\vspace{0.2cm}

Kagome lattice materials have attracted considerable attention due to their intriguing topological properties and potential applications in next-generation quantum and spintronic technologies. In particular, rare-earth permanent magnets with Kagome structure provide an ideal platform that combines robust magnetism with nontrivial quantum phenomena. However, their anomalous transport properties, particularly thermoelectric responses, remain insufficiently explored. In this work, we perform systematic first-principles calculations on the anomalous Hall and anomalous Nernst effects in Kagome permanent magnets \textit{RCo}$_5$ (\textit{R}=Ce, La, Sm, Gd). We find that CeCo$_5$ exhibits a pronounced anomalous Hall conductivity of $\sim 1500\ \Omega^{-1}\mathrm{cm}^{-1}$ while GdCo$_5$ displays a substantial anomalous Nernst conductivity of $11\ \mathrm{A\!\cdot\!m^{-1}\!\cdot\!K^{-1}}$ within $\pm 0.1$ eV of the Fermi energy, both comparable to or surpassing the measured intrinsic values reported in many typical Weyl and Heusler magnets. These exceptional anomalous transport properties originate from Berry curvature hotspots near spin-orbital coupling induced band gaps. If validated, these theoretical predictions would be important for Berry-curvature-driven transport in magnetic intermetallic. Our results establish \textit{RCo}$_5$ compounds as versatile platforms for exploring Berry curvature-driven transport in tunable magnetic topological materials.

\vspace{0.8cm}

\noindent{\Large \bfseries I.\hspace{1.5cm}Introduction}

\vspace{0.2cm}

The rare-earth permanent magnets are generally referred to intermetallic compounds composed of rare-earth and transition-metal elements, such as NdFeB, SmFeN. These materials exhibit high coercivity and play a vital role in modern industries, including aerospace and microwave communications [1]. Kagome lattice materials, characterized by geometrically frustrated networks, have attracted considerable interest due to their abundant quantum phenomena, including spin liquid states [2--4] and fractional quantum Hall effect [5--7]. In addition, the Kagome lattice hosts flat bands and van Hove singularities(vHS) [6,8] which have been associated with unconventional superconductivity and charge density wave (CDW) [9--12]. When permanent magnets adopt a Kagome lattice structure, they simultaneously possess both high magnetic hardness and nontrivial topological properties [13--15]. Among the fertile Kagome materials, compounds such as \textit{RCo}$_5$(\textit{R}=rare earth) crystallize in the CaCu$_5$-type structure (space group P6/mmm), which can be regarded as an alternating stack of Kagome Co planes and rare-earth atomic layers stacked along \textit{c}-axis. These compounds represent a typical class of permanent magnets with Kagome lattice.

The anomalous Hall effect (AHE), in contrast to the ordinary Hall effect, refers to a transverse current that arises in magnetic systems due to intrinsic magnetic moments and spin-orbit coupling (SOC) [16,17]. Analogously, when an electric field is replaced by a temperature gradient, a transverse voltage drop can occur, known as the anomalous Nernst effect (ANE) [18]. Both AHE and ANE can originate from two distinct mechanisms: an intrinsic contribution, which is fundamentally related to Berry curvature (BC) of the electronic band structure and an extrinsic contribution, attributed to scattering processes such as skew scattering and side-jump mechanisms [19]. The intrinsic AHE, in particular, is directly linked to the anomalous velocity induced by the Berry phase, making the BC a central quantity in understanding topological transport phenomena [20,21].

Despite the growing understanding of Berry curvature-driven transport in various materials, most previous studies on Kagome lattice permanent magnets, namely \textit{RCo}$_5$ family, have primarily focused on their magnetic properties [22--29]. Only a few investigations of AHE and ANE in this system were reported while many members of this family are still lack of study. For example, Ngo calculated AHE in GdCo$_5$ [30], while Asuka investigated the anomalous Ettingshausen and Nernst effects in SmCo$_5$ and (Sm,Gd)Co$_5$ alloys, initiating the concept of ``thermoelectric permanent magnets'' by linking magnetic materials to thermoelectric performance [31,32]. In contrast, magnetic topological semimetals---such as GdPtBi [33], Co$_3$Sn$_2$S$_2$ [34], Mn$_3$Sn [35], and Co$_2$MnGa [18]---have been widely recognized as model systems for exploring large AHE and ANE responses. In these systems, both effects are understood to arises from pronounced Berry curvature near the Fermi level, often associated with SOC-induced gaps or symmetry-protected nodal lines. Notably, when mirror symmetry is broken in certain Heusler compounds, SOC can open a gap along nodal lines, leading to enhanced Berry curvature and strong intrinsic transport responses [36,37].

In this work, we present theoretical research on Kagome permanent magnets \textit{RCo}$_5$ (\textit{R}=Ce, La, Sm, Gd), focusing on their AHE and ANE. These four members are selected because they possess a fixed easy axis along the \textit{c}-axis, unlike compounds with \textit{R}=Pr, Nd, Dy, Ho, and Tb, which exhibit temperature-dependent spin reorientation transitions, with magnetization directions that vary across different temperature ranges. Our computational results reveal that all four members exhibit a large AHC within $\pm 0.1$ eV of the Fermi energy, with CeCo$_5$ showing the highest value. Its peak AHC reaches 1500 S/cm at 0.084 eV below the Fermi energy. To identify the origin of the giant AHC, we carefully study the BC of each member in the first Brillouin zone and find that the main contributions of BC in CeCo$_5$ stem from the $k_z$=0.2 and $k_z$=0.4 planes. In these planes the BC is strongly concentrated on energy gaps opened by spin-orbital coupling which contributes significantly to the AHC. Analogously, GdCo$_5$ exhibits the highest ANC among the four compounds within $\pm 0.1$ eV, reaching $11\ \mathrm{A\!\cdot\!m^{-1}\!\cdot\!K^{-1}}$ at approximately 80 meV above the Fermi energy and the main contribution from BC also stems from the BC located at gaps opened by the SOC. Our work demonstrates the coexistence of pronounced anomalous Hall and anomalous Nernst effects within the same compound family-namely, Kagome permanent magnets-offering a valuable platform for studying anomalous transport phenomena.

\section*{II.\hspace{1cm}Methods}

\subsection*{A.\hspace{0.5cm}Calculation of anomalous transport properties}

The Berry curvature in the system is calculated by the Kubo formula [16,19,38]

\begin{equation}
\Omega_{ij}^{n}
=
i \sum_{m\neq n}
\frac{
\langle n|\frac{\partial H}{\partial k_i}|m\rangle
\langle m|\frac{\partial H}{\partial k_j}|n\rangle
-
(i \leftrightarrow j)
}
{(E_n-E_m)^2}
\end{equation}

where $|\psi\rangle$, $E_n$ are eigenstate and eigenvalue of $H$, respectively. Given the Berry curvature, we can calculate the AHC via the Kubo formula [39]

\begin{equation}
\sigma_{ij}
=
\frac{e^2}{\hbar}
\sum_n
\int
\frac{d^3k}{(2\pi)^3}
\Omega_{ij}^{n} f_n
\end{equation}

The ANC proposed by Xiao [40] can be obtained by

\begin{equation}
\alpha_{ij}
=
-\frac{1}{T}\frac{e}{\hbar}
\sum_n
\int
\frac{d^3k}{(2\pi)^3}
\Omega_{ij}^{n}
\left[
(E_n-E_F)f_n
+
k_B T
\ln
\left(
1+\exp
\left(
\frac{E_n-E_F}{-k_B T}
\right)
\right)
\right]
\end{equation}

here, $T$ and $f_n$ are the temperature and Fermi-Dirac distribution function, respectively.  
AHC and ANC can both be formulated within a unified theoretical framework as [37]:

\begin{equation}
\lambda_{ij}
=
\frac{e^2}{\hbar}
\int
\frac{d^3k}{(2\pi)^3}
\Omega_{ij}^{n}
w_{\lambda}(E_n-E_F),
\qquad
\lambda=\sigma,\alpha
\end{equation}

where $w_\sigma$ and $w_\alpha$ are

\begin{equation}
w_\sigma(E)=f_n^T(E)
\end{equation}

\begin{equation}
w_\alpha(E)
=
-\frac{1}{eT}
\left[
E f_n^T
+
k_B T
\ln
\left(
1+\exp
\left(
\frac{E}{-k_B T}
\right)
\right)
\right]
\end{equation}

here $f_n^T$ is the Fermi-Dirac function at temperature $T$.

\subsection*{B.\hspace{0.5cm}First-principles calculations}

Electronic structure calculations were performed within the framework of density-functional theory (DFT), as implemented in Vienna Ab initio Simulation Package (VASP) [41]. The projector augmented wave (PAW) method was used to treat ionic potential, with the PAW\_PBE datasets (POTCAR tags) Co (06Sep2000), Ce(04Sep2000), La(06Sep2000), Sm(17Dec2003), Gd(23Dec2003). The exchange-correlation potential was treated using the Perdew-Burke-Ernzerhof (PBE) form of the generalized gradient approximation (GGA) [42], with spin-orbital coupling included in all calculations in a fully noncollinear setup, and the spin quantization axis along [001].

The plane-wave cutoff of 500 eV was applied and $\Gamma$-centered Monkhorst-Pack mesh of $14\times14\times18$ was used for Brillouin-zone sampling. The convergence criterion for electronic self-consistency was set to $10^{-6}$ eV and Hellmann-Feynman forces were converged to less than $0.02$ eV/\AA. For the smearing scheme, the tetrahedron method with Bl\"ochl corrections (ISMEAR = $-5$) was employed in all calculations to ensure accurate total energies and density of states.

Magnetic initial conditions were assigned according to the rare-earth and transition-metal sublattices, e.g., $\sim5\ \mu_B$ for Ce/La/Sm/Gd atoms and $\sim2\ \mu_B$ for Co atoms. Experimental crystal structures were adopted as reported in Refs. [24,28,43]. The values of the effective Hubbard $U_{\rm eff}$ (Hubbard $U$ minus Hund’s coupling $J$ [44]) used in the calculations were $5.0$ eV for Ce and La, and $6.0$ eV for Sm and Gd, based on literature values [45--47] and were further tested to ensure reasonable agreement between the calculated and experimental magnetic moments [26].

The rare-earth $4f$ states were explicitly treated as valence electrons within the DFT+$U$ framework. Benchmark tests comparing the valence and open-core treatments are provided in the supplementary materials [48].

Based on the electronic structure, a tight-binding Hamiltonian was constructed using the maximally localized Wannier function (MLWF) method [49], as implemented in the Wannier90 package [50]. The $d,f$ orbitals of $R$ atoms and the $s,p,d$ orbitals of Co atoms were selected as projectors for wannierization procedure.

Based on the wannierized tight-binding Hamiltonians, the AHC and ANC were calculated using the Kubo formula based on Eq. 2 and Eq. 3, as implemented in Wanniertools package [51] with a Lorentzian broadening parameter $\eta=0.001$ eV at $T=300$ K. The Berry curvature was integrated over $\Gamma$-centered meshes up to $101\times101\times101$.

To benchmark the reliability of our calculations, we extracted the spin and orbital contributions to the magnetic moments on the rare-earth and Co sites. The results, summarized in Table I, show that the calculated total magnetic moments per formula unit are in reasonable agreement with experimental values, validating the calculation setups.

\vspace{0.3cm}

\textbf{Table I. Calculated magnetic moments of $RCo_5$ and experimental values. All quantities are in $\mu_B$.}

\begin{center}
\begin{tabular}{c c c c c}
\hline
 & Re moment ($\mu_B$) & Co moment ($\mu_B$) & Total moment & Exp. ($\mu_B$/f.u.) \\
 & (spin/orbital/total) & (spin/orbital/total) & ($\mu_B$/f.u.) & \\
\hline
CeCo$_5$ & -0.787/0.307/-0.480 & 7.109/0.662/7.771 & 7.291 & $\sim7.1$ [52,53] \\
LaCo$_5$ & -0.206/0.023/-0.183 & 7.499/0.703/8.202 & 8.019 & $\sim8.1$ [54,55] \\
SmCo$_5$ & 5.132/-0.569/4.563 & 7.552/0.664/8.216 & 12.779 & $\sim12.15$ [28] \\
 & & & & $\sim12.86$ [56] \\
GdCo$_5$ & -7.296/-0.007/-7.303 & 7.814/0.607/8.421 & 1.118 & $\sim1.37$ [57] \\
\hline
\end{tabular}
\end{center}

\section*{III.\hspace{1cm}Results and discussion}

\subsection*{A.\hspace{0.5cm}Basic electronic structure of $RCo_5$}

\begin{figure*}[htbp]
\centering
\includegraphics[width=\textwidth]{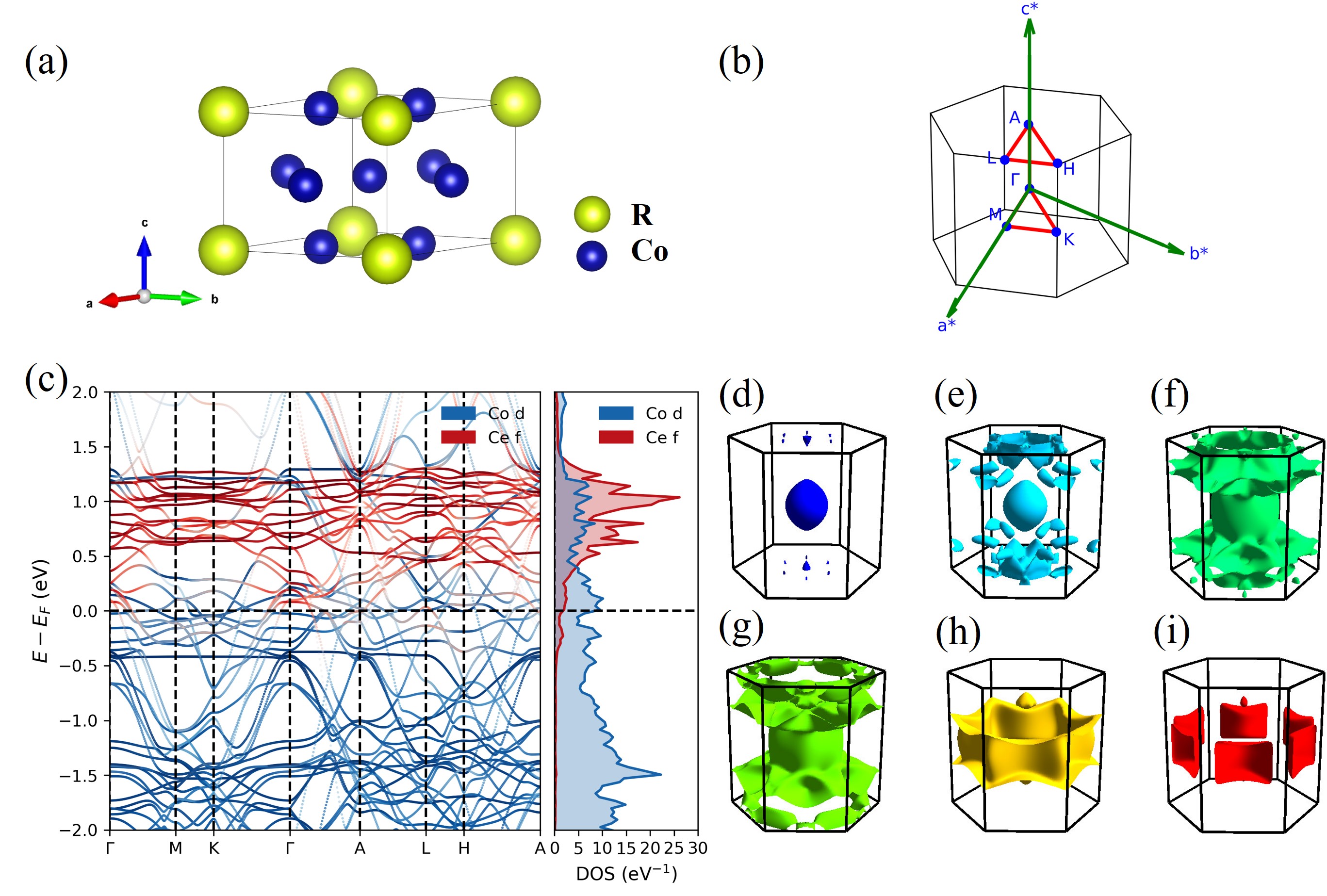}
\caption{Crystal and electronic structure of $RCo_5$. 
(a) Crystal structure of $RCo_5$ compounds constructed with the VESTA program [60]. 
(b) The first Brillouin zone and high-symmetry lines. 
(c) Spin-summed band structure and density of states of CeCo$_5$, calculated with spin-orbit coupling (SOC). 
(d)-(i) present the three-dimensional Fermi surfaces of the six bands crossing the Fermi level in CeCo$_5$.}
\label{fig1}
\end{figure*}

We first present the crystal structure and basic electronic structure of $RCo_5$ in Fig.~\ref{fig1}. 
$RCo_5$ family crystallizes in a stacked $P6/mmm$ (No.~191) structure. 
$R$ atoms occupy the center of hexagons in $R$-Co1 layers and Co2 layers form the Kagome lattice as shown in Fig.~1(a). 
The spin-summed band structure and density of states of CeCo$_5$, calculated with SOC are shown in the Fig.~1(c). 
The density of states reveals that the $d$-orbital component dominates near the Fermi level in CeCo$_5$, while the $f$-orbitals also contribute, primarily appearing above the Fermi level. 
(d)-(i) are the three-dimensional Fermi surface of the six individual bands crossing the Fermi level for CeCo$_5$. 
The other members’ Fermi surfaces and band structures are shown in the supplementary materials [48].

In each branch of the Fermi surface of CeCo$_5$ as shown in (d)-(i), the weight of five $d$ components is relatively uniform, while seven $f$ components account for a large proportion in the fifth and sixth branches, contributing to the $f$ states at the Fermi level in the DOS. 
From the band structure of the four members (see the supplementary materials [48]), we see that two sets of flat bands which are composed of Ce $f$ orbitals located at around 0.6 eV above the Fermi level and Co $d$ orbitals pinned at 0.4 eV below the Fermi level. 
The flat bands in $RCo_5$ differ significantly from those in $AV_3Sb_5$ compounds, where the flat bands, composed of $d$ orbitals, are located above the Fermi level [58,59].

\subsection*{B.\hspace{0.5cm}Anomalous Hall effect in RCo5}

\begin{figure*}[htbp]
\centering
\includegraphics[width=\textwidth]{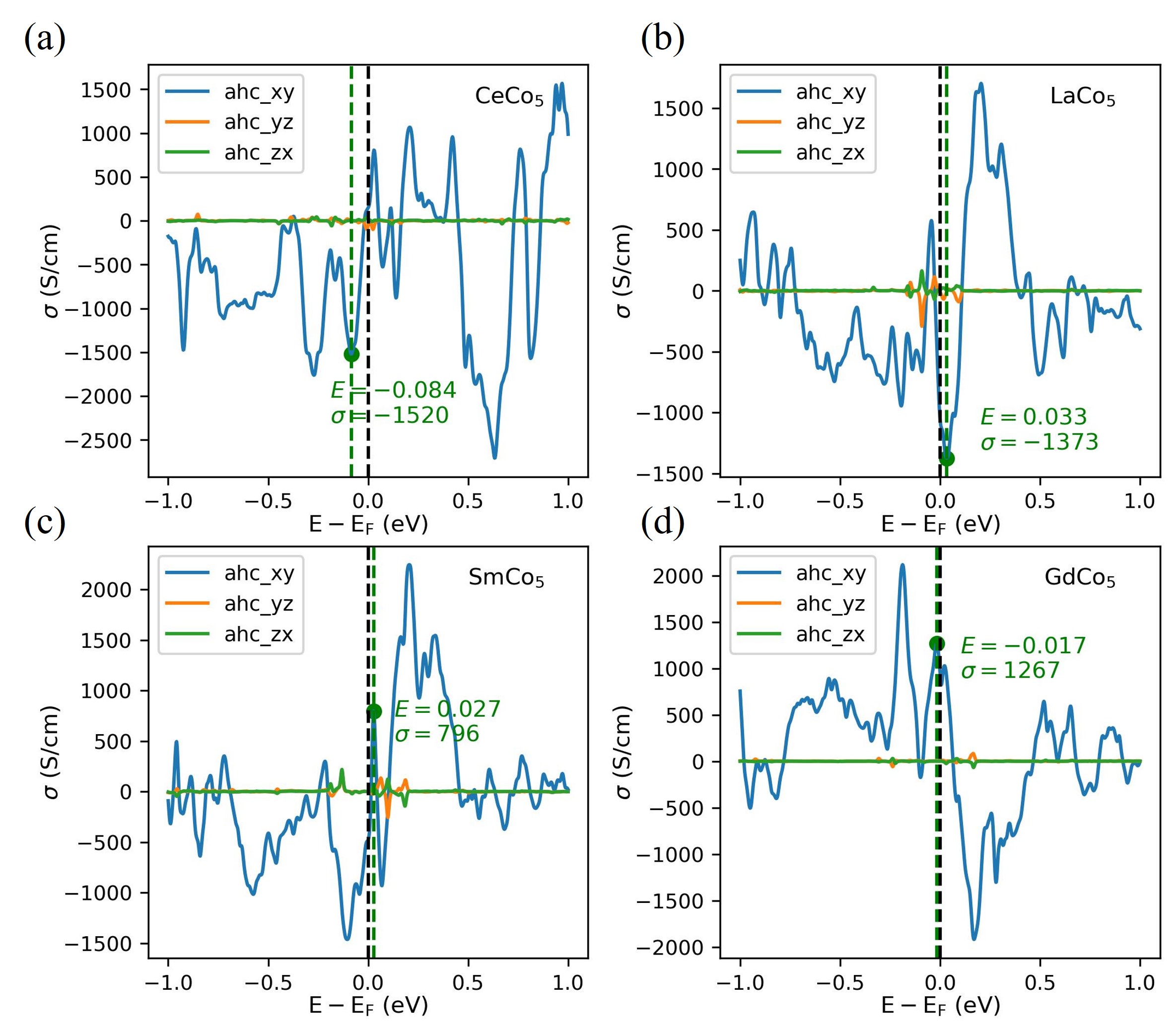}
\caption{Calculated anomalous Hall conductivity (AHC) for CeCo$_5$ (a), 
LaCo$_5$ (b), SmCo$_5$ (c), and GdCo$_5$ (d), respectively. The green lines 
highlight the maximum AHC values within $\pm 0.1$ eV of the Fermi level 
for each compound, with the corresponding energy positions and magnitudes 
indicated by green circles.}
\label{fig2}
\end{figure*}

Fig.~\ref{fig2}(a)-(d) are calculated anomalous Hall conductivities of CeCo$_5$, LaCo$_5$, SmCo$_5$, GdCo$_5$, respectively. 
The CeCo$_5$, LaCo$_5$, SmCo$_5$ exhibit ferromagnetic ordering [24,28,61], whereas GdCo$_5$ adopts a ferrimagnetic configuration, in which Gd $4f$ moments are antiparallel to the Co $3d$ moments [23]. 
Here we utilize the experimental lattice parameters [24,28,43] and adopt different Hubbard $U$ values from the literature [45--47] to fit the experimental magnetic moment [26]. 
From the calculated results, it is evident that all compounds exhibit giant values of AHC in the vicinity of the Fermi level, with CeCo$_5$ showing the largest magnitude reaching almost $1500~\Omega^{-1}\text{cm}^{-1}$.

In the calculations, all compounds were treated in collinear magnetic states with the spin quantization axis strictly along the crystallographic $c$ axis. 
Specifically, LaCo$_5$, CeCo$_5$, and SmCo$_5$ were modeled as ferromagnets with all moments parallel to $c$, while GdCo$_5$ was modeled as a ferrimagnet and the net magnetization still aligned along $c$. 
Because the magnetization is strictly aligned $c$, the anomalous Hall and Nernst tensors reduce to an axial form with only in-plane antisymmetric components being finite, while all out-of-plane components vanish by symmetry.

\begin{figure*}[htbp]
\centering
\includegraphics[width=\textwidth]{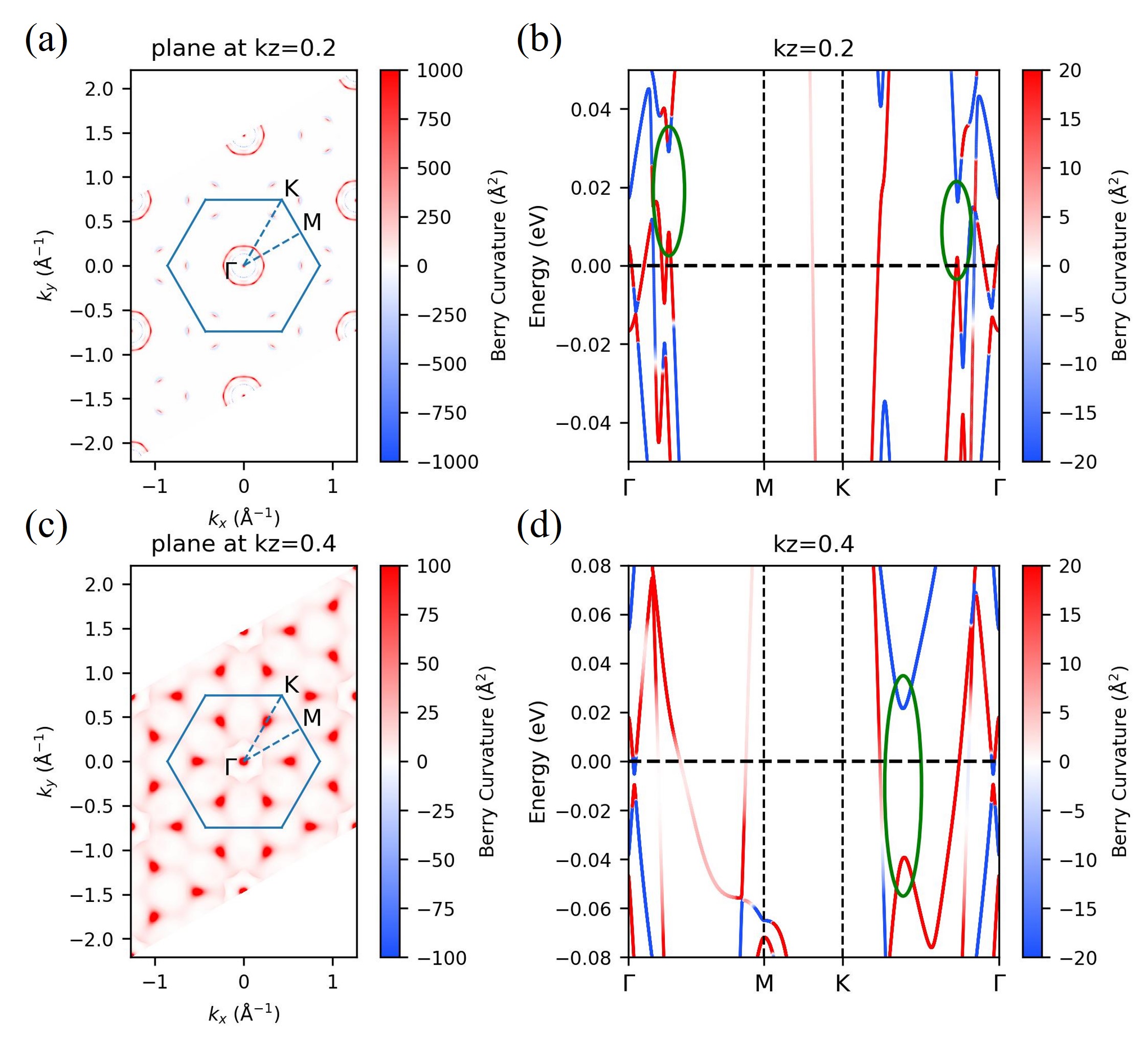}
\caption{Berry curvature analyses of CeCo$_5$. (a, c) The Berry curvature 
distributions on the $k_z=0.2$ and $k_z=0.4$ planes of the Brillouin zone, 
respectively. Panels (b, d) are the corresponding band structures along 
the $\Gamma$–M–K–$\Gamma$ path colored by the values of the Berry 
curvature. The positions marked by green ellipses correspond to the Berry 
curvature hotspots in panels (a) and (c).}
\label{fig3}
\end{figure*}

As illustrated in Fig.~\ref{fig3}(a)-(d), CeCo$_5$—which exhibits the highest AHC—is selected as a representative compound for investigating the underlying mechanism. 
From previous research, we already know that AHC is directly related to Berry curvature and proportional to the sum of its integral over all occupied states below the Fermi level in the momentum space [16]. 
To better understand the underlying mechanism, we evaluate the BC distribution summed over all occupied states below the Fermi level on different $k_z$ planes ranging from $k_z=0$ to $k_z=0.5$ in the first Brillouin zone (all $k_z$ values are in the unit of $2\pi/c$). 
This allows us to clearly identify which part of the band structure contribute most significantly to the BC and further contribute to the AHC. 
We find that the BC is most pronounced on $k_z=0.2$ and $k_z=0.4$ planes, as shown in Fig.~\ref{fig3}(a) and (c). 
The corresponding band structure on these planes are presented in Fig.~\ref{fig3}(b) and (d), respectively.

As shown in Fig.~\ref{fig3}(a), a ring-shaped region with strongly enhanced positive BC is observed around the $\Gamma$ point in the first Brillouin zone (outlined by the blue solid line). 
This annular distribution resembles a gapped nodal line structure, but further confirmation is required, as will be discussed in the following section. 
In the corresponding band structure in Fig.~\ref{fig3}(b), two band gaps can be observed along the $\Gamma$–M and $\Gamma$–K directions (highlighted by green circles), with the Fermi level lying within these gaps. 
Notably, the band located below the Fermi level contributes a positive BC, consistent with the ring-shaped distribution seen in Fig.~\ref{fig3}(a).

On the $k_z=0.4$ plane, the BC distribution exhibits a pattern similar to that on $k_z=0.2$ plane. 
As shown in Fig.~\ref{fig3}(c), a pronounced peak of BC appears along the $\Gamma$–K path close to the K point within the first Brillouin zone. 
In the corresponding band structure (Fig.~\ref{fig3}(d)), a band gap is observed along the same $\Gamma$–K direction. 
The lower band at the gap position (i.e., below the Fermi level) contributes a positive BC (highlighted by the green circle), which corresponds to the BC hotspot near the $\Gamma$–K path in the BC distribution at $k_z=0.4$.

\begin{figure*}[htbp]
\centering
\includegraphics[width=\textwidth]{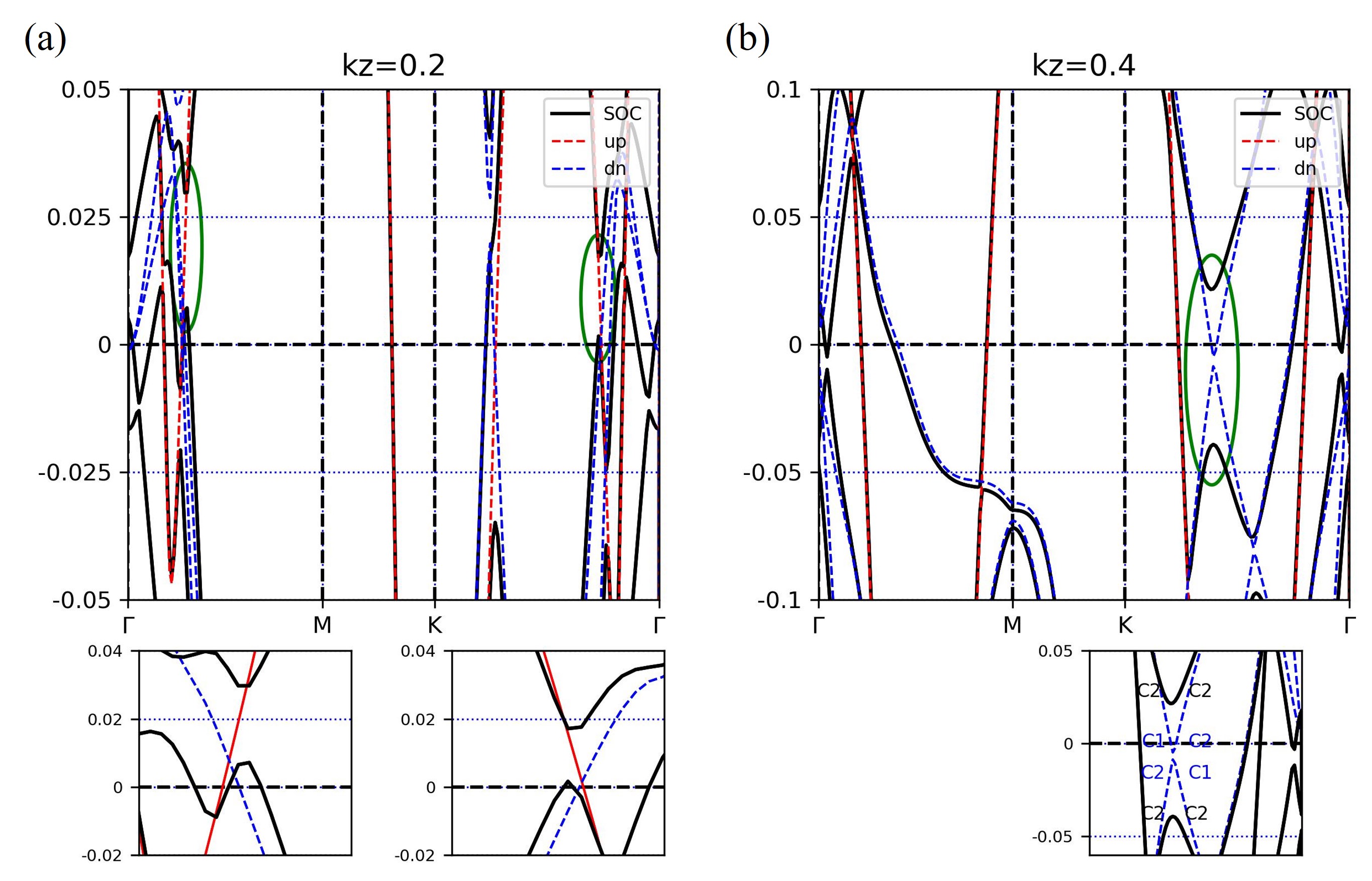}
\caption{Band structure and symmetry analysis. (a) and (b) are the band 
structures with $k_z=0.2$ and $k_z=0.4$, respectively. The lower row 
shows zoomed-in views of the regions enclosed by green ellipses in the 
upper panel. The lower panel of (b) also shows the irreducible 
representations of the bands.}
\label{fig4}
\end{figure*}

To investigate the origin of the pronounced BC near the gap observed in Fig.~\ref{fig3}, we analyzed the band structure on the corresponding $k_z$ planes, as shown in Fig.~\ref{fig4}(a)-(b). 
On the $k_z=0.2$ plane, when SOC is taken into account, the crossing of spin-up and spin-down bands opens a band gap. 
This band crossing differs from that observed in Heusler compounds Fe$_2$Mn$X$ [36] and Co$_3$Sn$_2$S$_2$ [34], as it does not involve crossing within the same spin channel. 
Therefore, once the SOC opens the gap, no gapped nodal lines are formed as in Heusler compounds.

On the $k_z=0.4$ plane, the band crossing is symmetry-protected in the absence of SOC. 
However, once SOC is introduced and a gap opens, the two bands forming the gap are no longer symmetry-protected, as confirmed by analyzing their irreducible representations. 
Actually, in regions where SOC induces band splitting, the Berry curvature becomes sharply localized near the resulting band gaps. 
This reflects one of the key intrinsic mechanisms underlying the giant AHE.

\begin{figure*}[htbp]
\centering
\includegraphics[width=\textwidth]{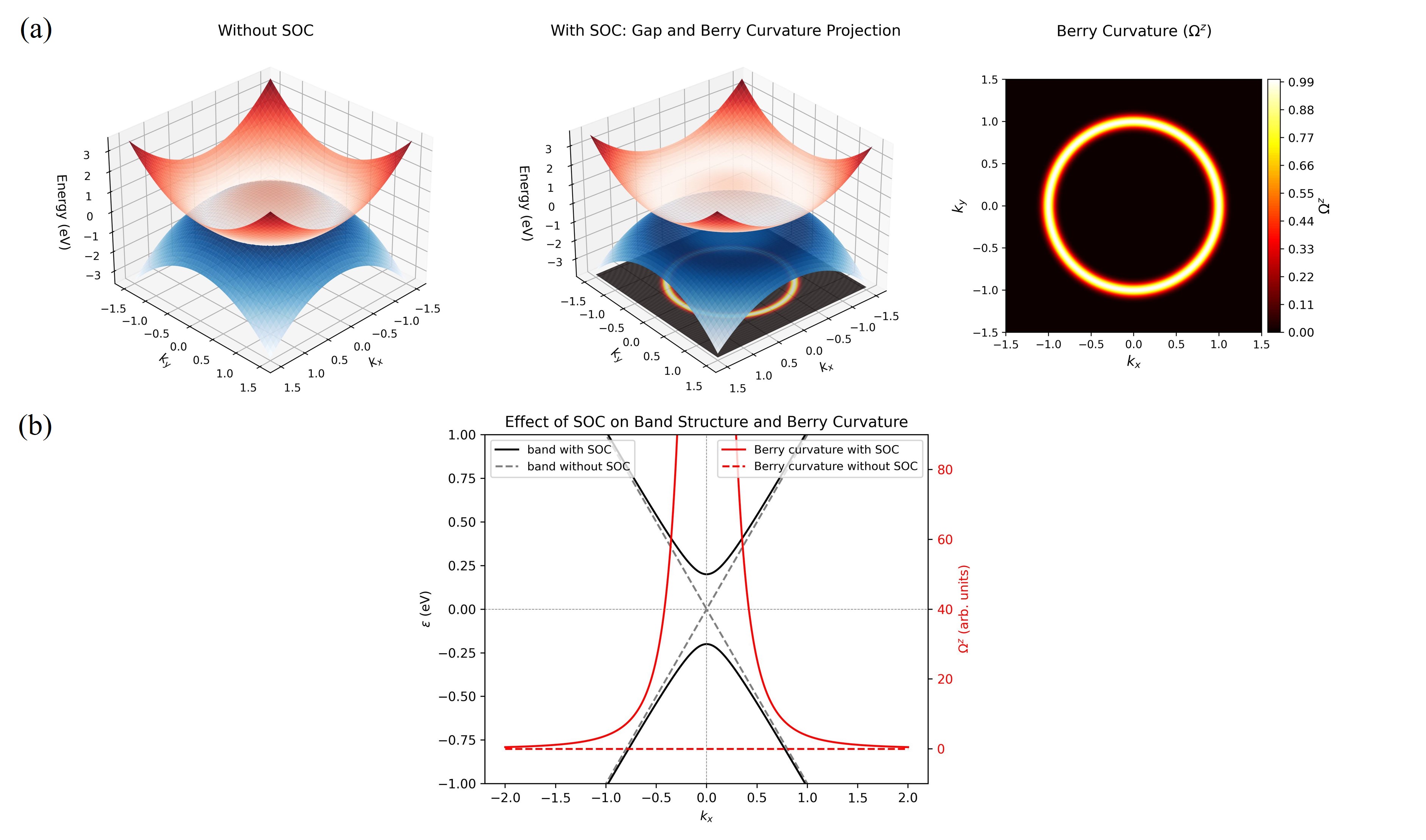}
\caption{Schematic models illustrating the origin of the Berry curvature 
from SOC-induced band gaps. (a) Band inversion forms a nodal ring without 
SOC (left panel), with vanishing Berry curvature across the momentum 
space (middle panel). Right panel: SOC opens a gap along the ring, 
generating a ring-like Berry curvature hotspot. (b) Two linear bands 
cross at a Dirac point. SOC lifts the degeneracy and induces a peaked 
Berry curvature centered at the gap. These two mechanisms correspond to 
the scenarios in Fig.~\ref{fig4}(a)–(b) and account for the Berry 
curvature distributions shown in Fig.~\ref{fig3}(a) and Fig.~\ref{fig3}(c), 
respectively.}
\label{fig5}
\end{figure*}

To provide an intuitive and quantitative understanding of the BC hotspots revealed in Fig.~\ref{fig3}, 
we constructed two minimal models that simulate the SOC-induced band gap scenarios depicted in 
Fig.~\ref{fig5}(a)-(b). These two cases correspond directly to the band structures analyzed in 
Fig.~\ref{fig4}(a) and Fig.~\ref{fig4}(b). Specifically, Fig.~\ref{fig5}(a) reflects the ring-like 
Berry curvature distribution associated with the SOC-induced gap on the $k_z=0.2$ plane shown in 
Fig.~\ref{fig3}(a), whose underlying band structure is discussed in Fig.~\ref{fig4}(a). Similarly, 
Fig.~\ref{fig5}(b) captures the Dirac-type Berry curvature peak corresponding to the SOC-induced 
gap near the K point on the $k_z=0.4$ plane, as analyzed in Fig.~\ref{fig3}(c) and band structure of 
Fig.~\ref{fig4}(b).

Fig.~\ref{fig5}(a) shows the schematic band structure without spin-orbital coupling. In this case, 
the conduction and valence bands invert and form a nodal ring in the $k_x$-$k_y$ plane. This nodal 
ring is protected by symmetry, and therefore, the Berry curvature remains strictly zero throughout 
the entire nodal ring. Upon introducing SOC, the degeneracy of the nodal ring is lifted, opening a 
gap along the ring. As a result, a ring-shaped distribution of sharply localized Berry curvature 
emerges near the band gap. This pattern closely resembles the annular Berry curvature hotspot 
observed in the results shown in Fig.~\ref{fig3}(a).

Fig.~\ref{fig5}(b) illustrates a scenario where two linearly crossing bands form a Dirac-like 
dispersion at a single point in momentum space. Without SOC, the crossing remains gapless and the 
Berry curvature is absent. When SOC is introduced, a band gap opens at the crossing point and the 
Berry curvature becomes sharply peaked in the vicinity of this gap. This is consistent with the 
sharply localized Berry curvature peak observed along the $\Gamma$-K path in Fig.~\ref{fig3}(c), 
which corresponds to the SOC-induced avoided crossing analyzed in Fig.~\ref{fig4}(b).

The Berry curvature of a two-band model can be formally written as [16]:

\begin{equation}
\Omega^{z}(\mathbf{k})=
\frac{1}{2}
\frac{\vec{h}\cdot
\left(
\partial_{k_x}\vec{h}\times\partial_{k_y}\vec{h}
\right)}
{|\vec{h}|^{3}}
\end{equation}

where $\vec{h}(\mathbf{k})=(h_x,h_y,h_z)$ are the momentum-dependent coefficients of the Pauli 
matrices in the two-band Hamiltonian $H(\mathbf{k})=h_x\sigma_x+h_y\sigma_y+h_z\sigma_z$. In our 
nodal ring model, $h_z(\mathbf{k})=k^2-k_0^2$ describes the band inversion that forms the ring, and 
SOC is introduced via a constant mass term $h_x=\Delta$. Assuming $h_y=0$, the Berry curvature 
simplifies to

\begin{equation}
\Omega^{z}(k)=
\frac{4\Delta r^{2}}
{\left[(r^{2}-k_0^{2})^{2}+\Delta^{2}\right]^{3/2}},
\end{equation}

where $r=\sqrt{k_x^{2}+k_y^{2}}$. This analytic expression demonstrates that the Berry curvature is 
strongly peaked around the ring defined by $r=k_0$, forming a Berry curvature hot ring consistent 
with our numerical findings.

To complement this, we also examine a simpler 1D Dirac-type model. In the absence of SOC, the band 
structure consists of linear crossings at $k=0$, and the Berry curvature vanishes. The energy 
dispersion in this case is given by

\begin{equation}
\varepsilon_{\pm}=\pm |k|,
\end{equation}

which describes the massless Dirac fermions. When SOC is introduced, it opens a gap at the Dirac 
point. This is typically modeled by adding a mass term $\Delta$, resulting in the massive Dirac 
Hamiltonian with energy eigenvalues

\begin{equation}
\varepsilon_{\pm}^{SOC}=\pm\sqrt{k^{2}+\Delta^{2}}.
\end{equation}

This dispersion reflects the lifting of degeneracy due to SOC.

The corresponding Berry curvature for the bands can be derived from the general two-band expression. 
For this model, the Berry curvature takes the form

\begin{equation}
\Omega^{z}(k)=
\frac{\Delta}{(k^{2}+\Delta^{2})^{3/2}}.
\end{equation}

This function is sharply localized around $k=0$ and decays rapidly for larger $k$, characterizing 
the intrinsic anomalous Hall effect arising from SOC-induced band gaps.

Together, these two models establish a direct link between SOC-induced gap openings and the formation 
of sharply localized Berry curvature. These insights provide an intuitive and quantitative 
understanding of the intrinsic mechanism underlying the AHE in the $RCo_5$ family, particularly in 
CeCo$_5$, where the band topology and SOC jointly give rise to the observed giant anomalous 
transport response. For a more quantitative analysis of the Berry curvature dependence on SOC and 
band gap size, see supplementary Fig.~S11 [48].

\subsection*{C.\hspace{0.5cm}Anomalous Nernst effect in $RCo_5$}

\begin{figure*}[htbp]
\centering
\includegraphics[width=\textwidth]{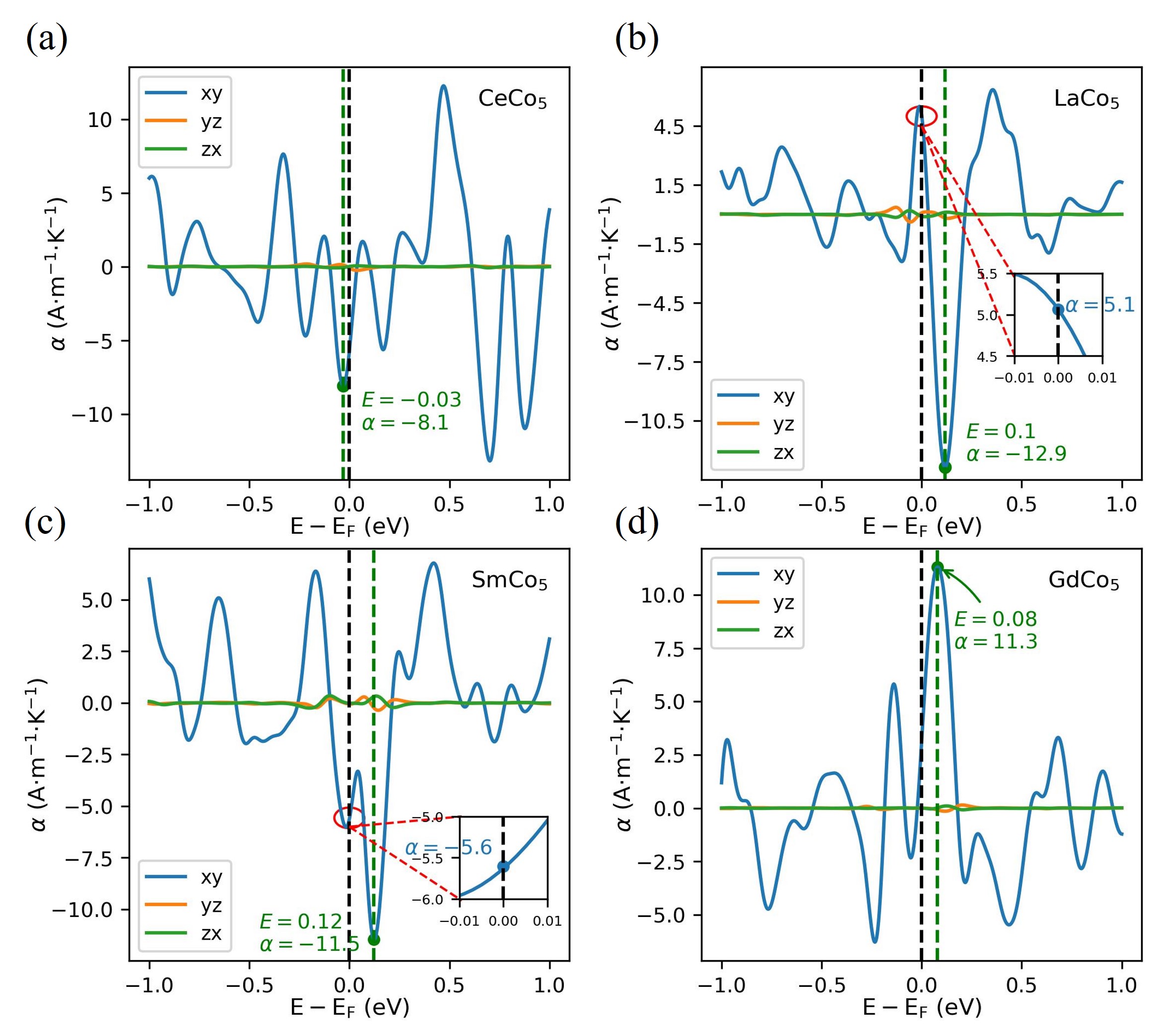}
\caption{Calculated anomalous Nernst conductivity (ANC) for CeCo$_5$ (a), 
LaCo$_5$ (b), SmCo$_5$ (c), and GdCo$_5$ (d). The green lines and dots 
indicate the maximum values within $\pm 0.1$ eV of the Fermi level. 
The insets of (b) and (c) show ANC values at the Fermi level.}
\label{fig6}
\end{figure*}

Fig.~\ref{fig6}(a)-(d) show the ANC of $RCo_5$ compounds at 300 K. 
For CeCo$_5$ and GdCo$_5$, the ANC reaches maximum values of 
$-8~\mathrm{A\cdot m^{-1}\cdot K^{-1}}$ at approximately 30 meV below the 
Fermi level and $11~\mathrm{A\cdot m^{-1}\cdot K^{-1}}$ at around 80 meV 
above the Fermi level, respectively. 
At the Fermi level, the ANC are 
$-5.5~\mathrm{A\cdot m^{-1}\cdot K^{-1}}$ for CeCo$_5$ and 
$3.3~\mathrm{A\cdot m^{-1}\cdot K^{-1}}$ for GdCo$_5$. 
In contrast, the maxima for LaCo$_5$ and SmCo$_5$ are located at 
0.11 eV and 0.12 eV above the Fermi level, as marked by the green dashed 
lines in Fig.~\ref{fig6}(b) and Fig.~\ref{fig6}(c). 

Since experimental measurements of ANC are available for LaCo$_5$ and 
SmCo$_5$, we indicate not only the energy positions of their maximum peaks 
but also the ANC values exactly at the Fermi level to enable direct 
comparison with experiment. For CeCo$_5$ and GdCo$_5$ where experimental 
ANC data are not yet reported, we provide both the Fermi-level values and 
the nearby maxima, so that our predictions can serve as a reference for 
future measurements. Notably, the calculated ANC for SmCo$_5$ at the Fermi 
level is approximately 
$5.6~\mathrm{A\cdot m^{-1}\cdot K^{-1}}$, which is in reasonable agreement 
with the experimental value of 
$4.5~\mathrm{A\cdot m^{-1}\cdot K^{-1}}$ [32], supporting the validity of 
our computational results.

As shown in Eq.~(3), the ANC also depends on the BC in momentum space. 
The explicit forms of Eq.~(2)–(6) are provided in the \textit{Methods} 
section and we summarize their physical meanings here for clarity. 
A comparison with the expression for the AHC in Eq.~(2) reveals that the 
key difference between AHC and ANC lies in the weights multiplied by the 
BC. If both of them are expressed in a unified form, they can be written 
as in Eq.~(4), where Eq.~(5) and Eq.~(6) define the corresponding weight 
functions for AHC and ANC, respectively.

From this unified formulation, it is evident that $w_\lambda$ serves as 
the integral weight distribution function for the Berry curvature in both 
AHC and ANC, with the distinctions between the weight functions 
illustrated in Fig.~S12. Specifically, $w_\sigma$ ensures that the Berry 
curvature of all occupied states below the Fermi level contributes to the 
intrinsic AHC. In contrast, $w_\alpha$ reaches its maximum at the Fermi 
level and decays exponentially away from it, with the rate of decay being 
temperature-dependent. Consequently, only the Berry curvature within a 
narrow energy window near the Fermi level predominantly contributes to 
the ANC. Moreover, the expression for ANC explicitly involves temperature 
$T$, which influences the weight function $w_\alpha$. At higher $T$, the 
weighting $w_\alpha$ gets broader, allowing electron states slightly above 
the Fermi level to contribute to the ANC as well.

\begin{figure*}[htbp]
\centering
\includegraphics[width=0.92\textwidth]{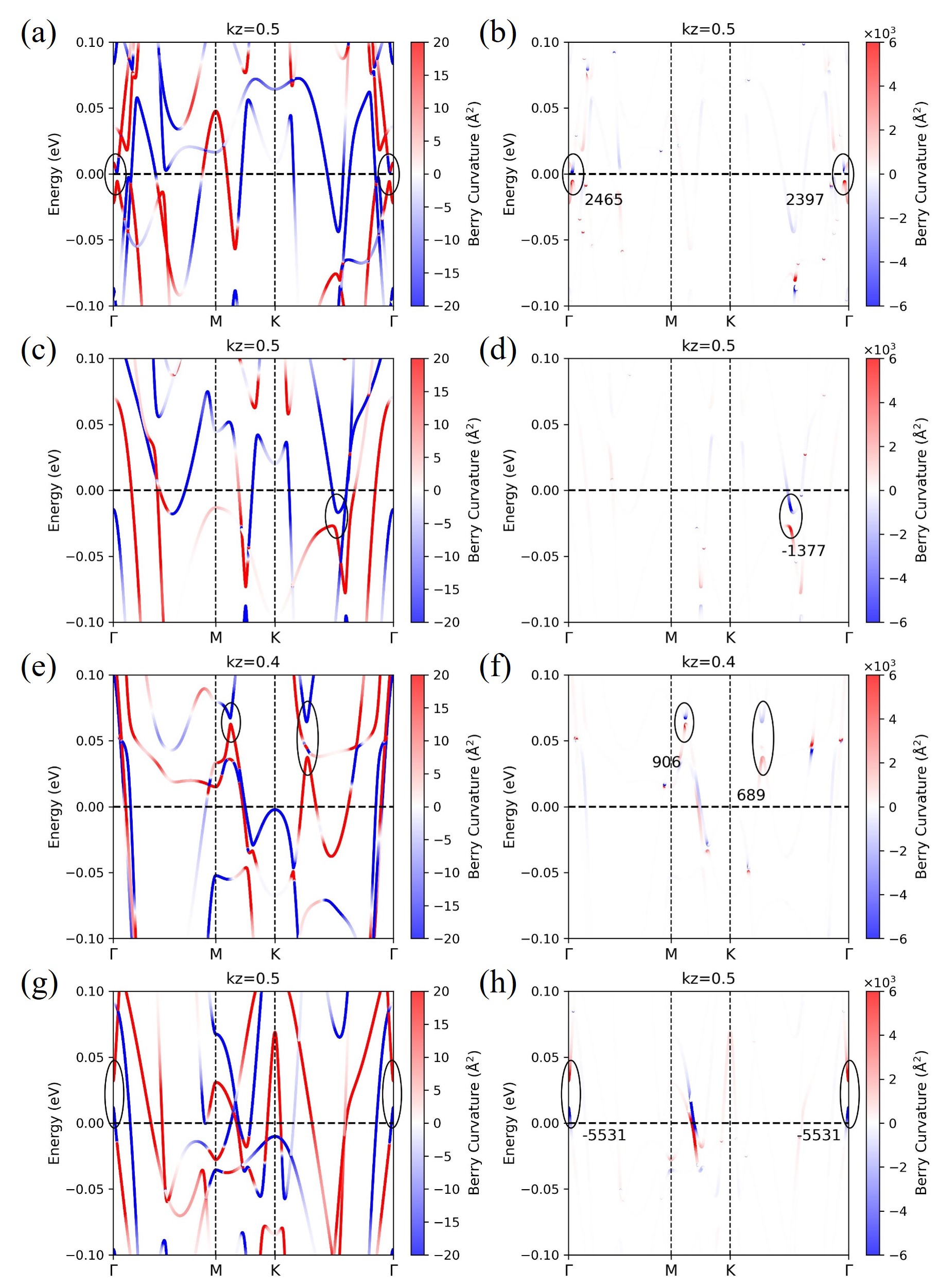}
\caption{Band-resolved Berry curvature analysis of $RCo_5$ for ANC. 
Panels (a)–(b): CeCo$_5$ ($k_z=0.5$); (c)–(d): LaCo$_5$ ($k_z=0.5$); 
(e)–(f): SmCo$_5$ ($k_z=0.4$); and (g)–(h): GdCo$_5$ ($k_z=0.5$). 
Panels (a), (c), (e), and (g) display the band structures along the 
$\Gamma$–M–K–$\Gamma$ path, colored by the values of the Berry curvature 
with a narrow color scale to provide a clearer view of the overall 
distribution, while the right panels (b), (d), (f), and (h) use a broader 
color scale to emphasize regions with the largest values. Black ellipses 
indicate locations where the Berry curvature reaches global maxima within 
the range $k_z=0$ to $k_z=0.5$.}
\label{fig7}
\end{figure*}

Based on the results of AHC, we observe that large BC typically arises 
near gapped regions in the band structure. To clearly identify the band 
gaps that contribute significantly to the ANC, we multiply the BC by the 
weight function $w_\alpha$ in our calculation. By examining the band 
structure across different $k_z$ planes, we identify particular band gaps 
that give rise to significant BC as illustrated in Fig.~\ref{fig7}(a)-(h). 
This approach allows us to determine the key band structure features 
responsible for the ANC.

In Fig.~\ref{fig7}, panels (a)-(h) show the energy and momentum-resolved 
Berry curvature distributions projected onto the band structures of the 
four compounds. The left column ((a), (c), (e), (g)) displays the BC with 
a narrow color scale ($\pm20$) to clearly visualize the overall 
distribution across the entire band, while the right column 
((b), (d), (f), (h)) uses a wider scale ($\pm6\times10^{3}$) to highlight 
the band segments with the strongest BC contributions. The regions 
enclosed by black ellipses indicate the specific band gaps that give rise 
to significant BC near the Fermi level. Notably, these are also the global 
maxima of BC across all $k_z$ planes from $k_z=0$ to $k_z=0.5$.

The BC values for LaCo$_5$ and SmCo$_5$ are taken at their original Fermi 
level, while for CeCo$_5$ and GdCo$_5$, the Fermi level is set to the 
energy positions of maximum ANC values marked in Fig.~\ref{fig6}. 
For the four compounds, the net BC at these marked regions—obtained by 
summing the contributions from the two bands forming the gap—follows the 
order (in absolute value): 
GdCo$_5$ ($-5531$), CeCo$_5$ ($4862$), SmCo$_5$ ($1595$) and 
LaCo$_5$ ($-1376$).

According to Eq.~(3), the ANC is directly proportional to the BC, which 
accounts for the descending order of the calculated ANC values in 
Fig.~\ref{fig6}: GdCo$_5$ (11.3), CeCo$_5$ ($-8.1$), SmCo$_5$ ($-5.7$) and 
LaCo$_5$ (5.0). Moreover, the overall sign difference between the BC and 
ANC—originating from the minus sign in Eq.~(3)—is clearly manifested in 
our computational results, demonstrating the internal consistency of the 
theoretical framework.

\subsection*{D.\hspace{0.5cm}Quantitative comparison of AHC and ANC in different compounds}

\begin{figure*}[htbp]
\centering
\includegraphics[width=\textwidth]{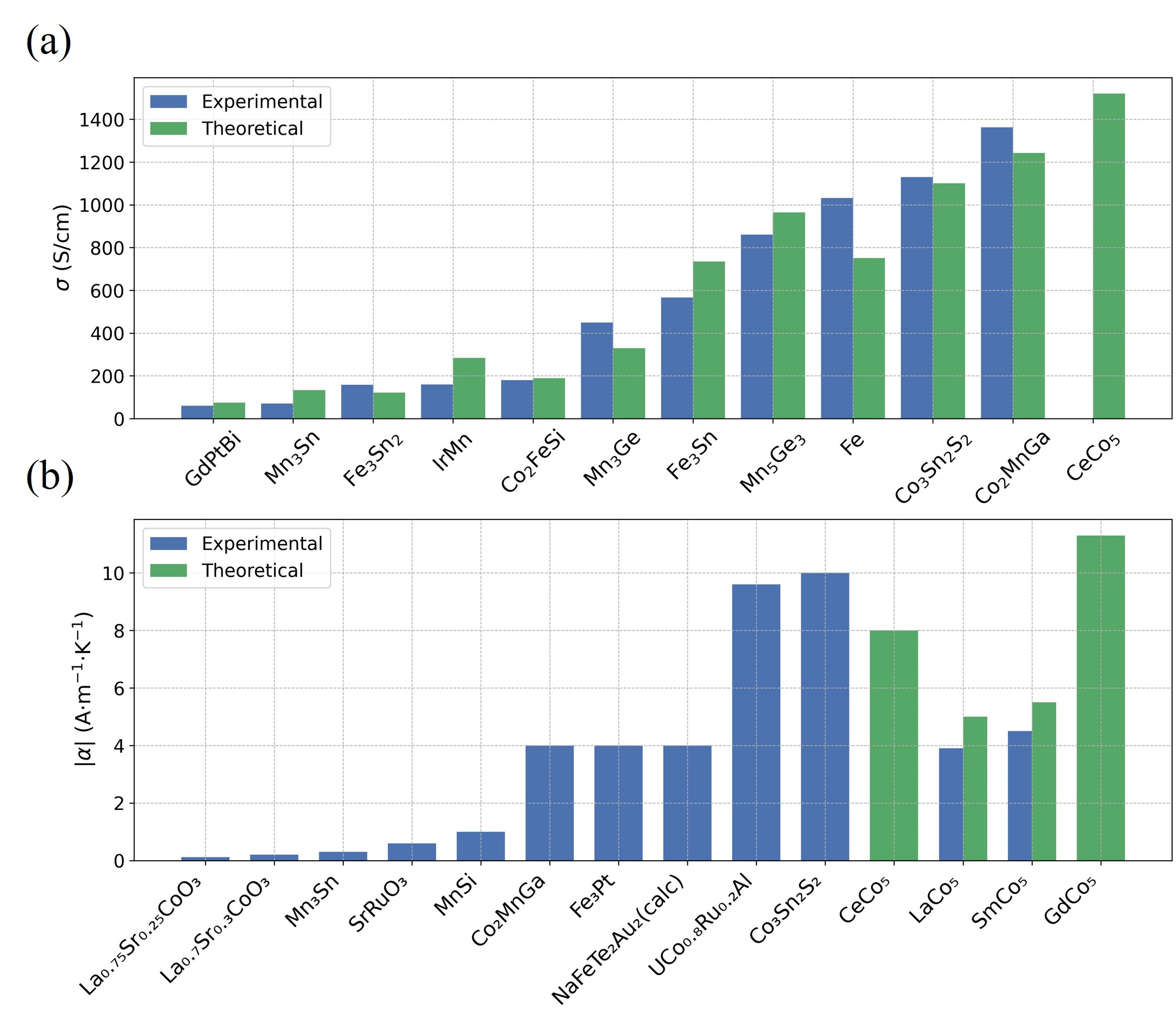}
\caption{Comparison of AHC and ANC across different materials. Comparison 
of our results and previously reported data for other materials. The 
reported data were taken from references [18,33–35,39,62–78] that can also 
be found in the supplementary materials [48]. Unless otherwise noted, the 
AHC values shown correspond to the intrinsic Berry-curvature contribution 
obtained either experimentally or theoretically.}
\label{fig8}
\end{figure*}

Fig.~\ref{fig8} presents a comparative analysis of the AHC and ANC across 
different compounds. Here we emphasize that the comparison in 
Fig.~\ref{fig8} is restricted to intrinsic contributions governed by 
Berry curvature either from experiments or from theoretical calculations. 
As shown in Fig.~\ref{fig8}(a), for most listed compounds, the theoretical 
and experimental values of intrinsic AHC exhibit reasonable agreement. 
This agreement indicates that the mechanism of intrinsic AHC governed by 
Berry curvature is consistent with experimental observations.

However, it should be noted that in some materials the experimentally 
measured AHC contains both extrinsic and intrinsic contributions, and the 
intrinsic components can differ substantially from the total value. 
For example, in KV$_3$Sb$_5$ [79], the extremely large AHC of 
$\sim15507$ S/cm originated from extrinsic mechanisms, namely skew 
scattering, rather than from the intrinsic Berry curvature. In Ref.~[80], 
the reported AHC of NdGaSi is about 1730 S/cm, while the intrinsic part 
amounts to $\sim1166$ S/cm. Similarly, the experimentally measured AHC of 
Mn$_{3.2}$Ge [81] is 1587 S/cm, whereas the calculated intrinsic value is 
only $\sim155.7$ S/cm.

We also note that the exceptionally large ANC reported for MnBi [82] 
($\sim44$ A$\cdot$m$^{-1}\cdot$K$^{-1}$ at 80 K) arises predominantly from 
extrinsic mechanisms such as magnon-drag, while at room temperature its 
ANC decreases to only $\sim2$--$3$ A$\cdot$m$^{-1}\cdot$K$^{-1}$. 
In YbMnBi$_2$ [78], an ANC of about 
10 A$\cdot$m$^{-1}\cdot$K$^{-1}$ is observed around 100 K. 
Therefore, when making such comparisons, it is essential to distinguish 
whether the experimental values correspond to intrinsic or extrinsic 
contributions.

Among the listed materials, CeCo$_5$ exhibits a pronounced theoretical 
AHC, on par with or exceeding that of the well-known Weyl semimetals such 
as Co$_3$Sn$_2$S$_2$ and Co$_2$MnGa. Considering the typical discrepancy 
between calculated and experimental values observed in these listed 
compounds, it is reasonable to expect that the experimental AHC of CeCo$_5$ 
should be slightly higher than, or at least comparable to those of 
Co$_3$Sn$_2$S$_2$ and Co$_2$MnGa.

Fig.~\ref{fig8}(b) compares the ANC values among these materials. 
GdCo$_5$ displays a substantial ANC within the $RCo_5$-type compounds, 
approaching or surpassing values observed in Co$_3$Sn$_2$S$_2$, 
Co$_2$MnGa, and numerous Heusler compounds reported in the literature.

Given that $RCo_5$ compounds are rare-earth permanent magnets with 
well-established fabrication processes, they offer excellent tunability 
via doping. For CeCo$_5$, density-of-states analysis shows that hole 
doping would shift the Fermi level to the energy where AHC reaches its 
maximum. Similarly, for GdCo$_5$, electron doping would shift the Fermi 
level to the energy at which the ANC reaches its maximum.

\section{Conclusions}

In summary, our first-principles predictions of the 
Kagome-structured rare-earth intermetallic compounds $RCo_5$ reveal that 
these materials exhibit exceptionally large Berry curvature-driven 
transport properties. In particular, CeCo$_5$ and GdCo$_5$ demonstrate 
notably large theoretical intrinsic AHC and ANC, respectively, comparable 
to or greater than those observed in many conventional ferromagnets and 
topological semimetals.

The theoretical intrinsic AHC of CeCo$_5$ is on par with or exceeds that 
of prototypical Weyl semimetals such as Co$_3$Sn$_2$S$_2$ and Co$_2$MnGa, 
while the ANC of GdCo$_5$ matches or exceeds values reported for numerous 
Heusler compounds in previous studies [83]. These predictions call for 
future experimental verifications.

These findings highlight $RCo_5$ compounds as a promising class of 
materials that simultaneously possess high structural stability, 
well-developed fabrication techniques, and substantial intrinsic 
transport coefficients. Moreover, the close agreement between theoretical 
and experimental AHC values across the $RCo_5$ family confirms that 
intrinsic Berry curvature mechanisms dominate in these systems.

Notably, the tunability of these transport properties via doping offers a 
practical route to optimize their performance for applications. The 
combination of high-coercivity permanent magnetism and topological 
transport phenomena in the $RCo_5$-type family offers a fertile platform 
for designing advanced spintronic and thermoelectric devices. These 
results motivate targeted experiments to verify the predicted optima. 
Future experimental studies, particularly on doped $RCo_5$, could further 
validate these findings and unlock their potential in next-generation 
devices.

\section*{Acknowledgements}

This work was supported by the Fundamental Research Funds for the Central 
Universities (Grant No.~2243300003), the Innovation Program for Quantum 
Science and Technology (Grant No.~2021ZD0302800), and the National 
Natural Science Foundation of China (Grant No.~12074041). 
The calculations were carried out on the high-performance computing 
cluster of Beijing Normal University in Zhuhai.

\section*{Data Availability}

The data that support the findings of this article are not publicly 
available upon publication because it is not technically feasible and/or 
the cost of preparing, depositing, and hosting the data would be 
prohibitive within the terms of this research project. 
The data are available from the authors upon reasonable request.

\newpage

\end{document}